\documentclass[prb,a4paper,showpacs,twocolumn,amsmath,amssymb]{revtex4}
\usepackage{graphicx}
\usepackage{color}

\makeatletter
\renewcommand{\@biblabel}[1]{(#1)}
\makeatother

\def\beq{\begin{equation}}
\def\ee{\end{equation}}
\def\bi{\begin {itemize}}
\def\ei{\end{itemize}}

\def\lsim
{\protect \raisebox{-0.75ex}[-1.5ex]{$\;\stackrel{<}{\sim}\;$}}
\def\gsim
{\protect \raisebox{-0.75ex}[-1.5ex]{$\;\stackrel{>}{\sim}\;$}}
\def\lsimeq
{\protect \raisebox{-0.75ex}[-1.5ex]{$\;\stackrel{<}{\simeq}\;$}}
\def\gsimeq
{\protect \raisebox{-0.75ex}[-1.5ex]{$\;\stackrel{>}{\simeq}\;$}}

\def\beq{\begin{equation}}
\def\ee{\end{equation}}

\def\bi{\begin {itemize}}
\def\ei{\end{itemize}}

\begin{document}

\title{Adhesion of microcapsules}
\author{Peter Graf, Reimar Finken, and Udo Seifert}

\affiliation{{II.} Institut f\"ur Theoretische Physik, Universit\"at Stuttgart,
  70550 Stuttgart, Germany}

\begin{abstract}
  The adhesion of microcapsules to an attractive contact potential is studied
  theoretically. The axisymmetric shape equations are solved numerically.
  Beyond a universal threshold strength of the potential, the contact radius
  increases like a square root of the strength. Scaling functions for the
  corresponding amplitudes are derived as a function of the elastic
  parameters.
\end{abstract}

\maketitle

\def\lsim
{\protect \raisebox{-0.75ex}[-1.5ex]{$\;\stackrel{<}{\sim}\;$}}

\def\gsim
{\protect \raisebox{-0.75ex}[-1.5ex]{$\;\stackrel{>}{\sim}\;$}}

\def\lsimeq
{\protect \raisebox{-0.75ex}[-1.5ex]{$\;\stackrel{<}{\simeq}\;$}}

\def\gsimeq
{\protect \raisebox{-0.75ex}[-1.5ex]{$\;\stackrel{>}{\simeq}\;$}}

{\sl Introduction. --} Microcapsules are hollow closed elastic capsules
experimentally prepared as layered polyelectrolyte sheets \cite{donath1998} or
through polymerization of surfactants coated on oil droplets immersed in
aqueous solution \cite{rehage2002}. Their elastic constants have been measured
by atomic force microscopy
\cite{fery2004,lulevich2004,mueller2005,elsner2006}, deformation in shear flow
\cite{walter2000} or osmotically induced swelling \cite{gao2001,vino2004}. For
potential applications not only the knowledge of their elastic constants is
crucial, but also, in particular, the understanding of their adhesive
properties to other capsules or membranes, to substrates
\cite{schwarz2000,nolte2004} or channel walls \cite{cordeiro2004} in
microfluidic devices. In a first systematic experimental study of adhesion to
glass surfaces using reflection interference contrast microscopy, adhesion
radii were measured as a function of capsule size and membrane thickness
\cite{elsner2004}. A systematic theory of such deformed shapes, however, is
lacking. The analysis of this experiment as well as previous theoretical
approaches are adaptions of the standard textbook treatment which involves
basically a scaling estimate of the deformation ignoring what are supposed to
be factors of order unity \cite{LL7}. Such a simple approach predicts that for
small deformations the adhesion radius scales as the square root of the
strength of the potential. On the other hand, in the related problem of the
adhesion of {\sl fluid} vesicles dominated by curvature elasticity, a
systematic solution based on solving numerically variational shape equations
has shown that adhesion requires a threshold strength of the potential
\cite{seifert1990}. A priori, there is no reason to believe that this
threshold vanishes if a finite shear elasticity is invoked in contrast to the
predictions based on the simple scaling picture.

\begin{figure}
\includegraphics[scale=0.25]{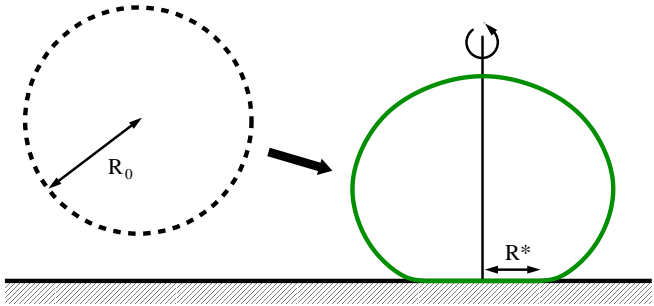}
\caption{Spherical microcapsule with radius $R_0$ undergoes an adhesion transition to an axisymmetric shape with adhesion radius $R^*$.}
\end{figure}
In this letter, we investigate systematically the adhesion of 
microcapsules in a contact potential by solving the corresponding
shape equations. We will restrict our treatment to axi-symmetric shapes, see Fig. 1. We therefore exclude configurations
where the adhesion area buckles inwards. Such configurations have been observed in
recent simulations where, however, the range of the potential was
comparable to the size of the capsules \cite{tamura2004,komura2005}. As a main qualitative result, we
indeed 
find that a finite universal 
threshold strength is required to induce adhesion beyond which
the adhesion radius scales like a square root in the excess strength.
The amplitude depends strongly on the elastic parameters.

{\sl Model. --} As an initial reference shape, we choose a sphere of radius
$R_0$. Any deformation of this sphere costs an elastic energy \beq F_s= \oint
dA \left( \frac{\lambda}{2} (u^l_l)^2 + \mu (u^i_k)^2 \right) \ee where
$\lambda$ and $\mu$ are the two-dimensional Lam\'e coefficients,
$K=\lambda+\mu$ is the compression modulus, and $u^i_k$ are the elements of
the strain tensor \cite{LL7,boal2000}. This relation is valid as long as
Hooke's law holds. The total energy of the capsule then reads \beq F= F_s +
\frac{\kappa}{2} \oint dA (2H-C_0)^2 - W A^*. \ee The second term is the
bending energy of the membrane where $H$ is the local mean curvature and
$\kappa$ the bending rigidity. Such a term is required to prevent sharp kinks
in an adhesion geometry. We also introduce a spontaneous curvature $C_0=2/R_0$
so that the original undeformed spherical capsule has no elastic energy
whatsoever. Generally a capsule can be adequately described by these thin shell
equations as long as the shell thickness is much smaller than the radius. This
assumption is generally fulfilled for capsules larger than a few microns. The
third term is the standard adhesion energy in a contact potential of strength
$W$ where $A^*\equiv \pi R^{*2}$ is the contact area.

The shape equations for this model can be derived by first parameterizing
an axisymmetric shape appropriately and then  setting the first
variation of $F$ to zero. This procedure can follow closely the
corresponding one for fluid vesicles with the additional feature that
for an elastic capsule tangential displacements have a physical significance
whereas for fluid vesicles they correspond to irrelevant reparameterizations 
of the shape and can thus be ignored \cite{seifert1997}. The technical details
will be published elsewhere.

The solutions of these shape equations and hence the ``phase diagram''
of adhesion depend on three dimensionless parameters
\beq
k\equiv \frac{K R_0^2}{\kappa},~~~m\equiv \frac{\mu R_0^2}{\kappa},~~~w\equiv \frac{W R_0^2}{\kappa} 
\ee
where $k$ and $m$ are the scaled compression and shear modulus, respectively,
whereas $w$ is the scaled adhesion energy. In this way, all energies are
referred to the scale of the bending energy. While typically $k,m \gg1$,
the scaled adhesion energy can easily be of order 1. Note that we have not
implied any volume constraint since at least for the polyelectrolyte shells
the membrane is supposed to be water permeable. Introducing such a volume
constraint, however,  would not pose any fundamental complication but would add
one more dimension to the phase diagram.

{\sl Adhesion radius. --}
By solving the shape equations numerically,
we find universally that the (scaled) adhesion radius of weakly adhered shapes
behaves as
\beq
r^*\equiv R^*/R_0\approx a(k,m)(w-w_c)^{1/2} .
\label{contact_radius}
\ee
The critical strength of adhesion $w_c=2$ is independent of the elastic
parameters and indeed the same as found for fluid vesicles without a volume
constraint \cite{seifert1990}. For a weaker contact potential, the potential energetic
 gain of an adhesion disc is smaller than the cost in curvature energy
to be paid. This balance is not modified by the elastic energies.

The dependence of the amplitude $a(k,m)$ on the elastic parameters is best 
discussed 
in several steps. For a capsule without shear elasticity, $m=0$, the
amplitude $a(k,0)$ decreases monotonically with $k$ from the value
\beq
a_0\equiv a(0,0)\simeq 0.58 ~~~~{\rm to} ~~~~
 a_\infty \equiv a(k\to \infty,0)\simeq 0.50.
\label{amplitudes}
\ee
For $k=0$, 
the presence of the non-zero spontaneous curvature energy stabilizes a 
finite size of the contact radius. For $k\to\infty$, the model 
corresponds to that of a fluid vesicle with area constraint. Both limit cases
can therefore be checked  independently using the shape equations of fluid
vesicles without and with area constraint, respectively. Note that without shear
elasticity, increasing the compression modulus from zero to infinity
thus leads to a decrease in the adhesion radius of only about 15 percent
for any given adhesion strength.

\begin{figure}
\includegraphics[scale=0.65]{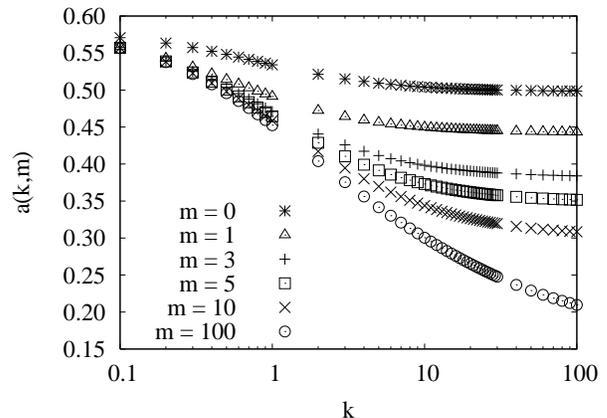}
\caption{Amplitude $a(k,m)$ of the adhesion radius (\ref{contact_radius}) as a function of the scaled compression modulus $k$ for various values of the scaled shear modulus $m$. Note that the abscissa is scaled logarithmically.}
\end{figure}
A quantitatively more dramatic effect arises for a non-zero shear modulus. In
Fig. 2, the amplitude $a(k,m)$ is shown as a function of $k$ for various
values of $m$. For fixed $k$, the amplitude decreases with increasing $m$
reaching a finite non-zero limit for $m\to \infty$. Similarly for fixed $m$ the amplitude
also decreases with increasing $k$ and reaches a non-zero value
in the limit $k \to\infty$. The limit $k\to\infty$ can quite well be fitted by
the power law \beq a(k \to \infty,m)\simeq a_\infty(1+m/m_c)^{-1/4}
\label{a_incompr}
\ee
with $a_\infty$ given above in eq. (\ref{amplitudes})
and $m_c\simeq 1.54$. 

These data for  general $k$ and $m$ 
almost collapse on a scaling plot
if these elastic constants are replaced by the 
(scaled) two-dimensional
Young modulus
\beq
y\equiv \frac{4km}{k+m}
\ee 
and the Poisson ratio
\beq
\sigma\equiv \frac{k-m}{k+m}.
\ee 
Fig. 3 reveals that the amplitude $a(y,\sigma)$ is almost
independent of the Poisson ratio as long as $y\gsim 6$, a
condition fulfilled if  ${\rm min}(k,m)\gsim 3$. For smaller
values of $y$, the amplitude becomes the larger the smaller $\sigma$.
Note that all curves have the same limit value $a(y \to 0,\sigma) \simeq 0.58$ except the curve for $\sigma=1$.
The two cases $k\to \infty$ and $m=0$ both correspond to $\sigma=1$. In the
first case the relation $y=4m$ holds and the amplitude can be calculated
directly from eq. (\ref{a_incompr}). In the second case, $y$ is always zero
and the amplitude varies from $a_0$ to $a_\infty$ for increasing $k$. 
\begin{figure}
\includegraphics[scale=0.65]{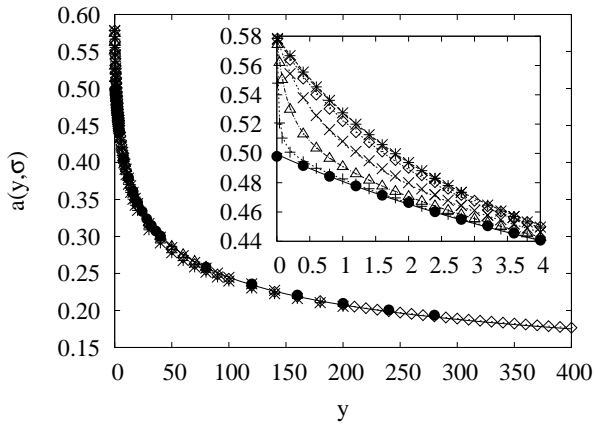}
\caption{Amplitude $a(y,\sigma)$ for different values of $\sigma$: $\sigma=-1$ ($*$), $\sigma=0$ ($\diamond$), $\sigma=0.6$ ($\times$),  $\sigma=0.9$ ($\bigtriangleup$), $\sigma=0.99$ ($+$), $\sigma=1$ ($\bullet$). For $y\gsim 6$ the plots almost collapse on a master curve. For $y\lsim 6$ a stronger dependence on $\sigma$ arises. The continuous lines are fits to the data points according to eq. (\ref{amplitude}).}
\end{figure}

For future reference, we extract from our data the fit
\beq
a(y,\sigma) \simeq \frac{a_\infty}{(1+y/y_c)^{1/4}} + \frac{a_0-a_\infty}{(1 + y/\hat y)^b}.
\label{amplitude}
\ee
with $\hat y = 1.85 (1-\sigma) + 5.58 (1-\sigma)^3$ and $b = 0.98 + 1.42(1-\sigma) + 2.71(1-\sigma)^3$.
The first term is based on the incompressible case (\ref{a_incompr}) discussed above with the crossover value $y_c \simeq 6.14$.
The deviations for small $y\leq 6$ are contained in the second term. 
This form provides a decent fit with a maximal deviation of 2
percent for all data shown in Fig. 3.

{\sl Comparison to experimental data. --}
Our theory can now be used to reanalyse experimental data. Recently Elsner et~al. \cite{elsner2004} investigated the adhesion of polyelectrolyte multilayer capsules (PMCs) on a glass surface. PMCs can be prepared with well-defined shell thickness, shell radius and surface energy. Therefore they are an ideal system to study the adhesive properties of microcapsules experimentally. The contact area and the form of the adhered capsules were reconstructed by reflection interference contrast microscopy. Given the two material properties Young modulus and Poisson ratio the contact potential can be derived by fitting data sets of the contact radius {\sl vs.} the capsule's wall thickness with fixed capsule's radius or the contact radius {\sl vs.} the capsule's radius with fixed capsule's wall thickness, respectively.

For thin isotropic shells of thickness $h$ and three-dimensional Young modulus $Y_3$ the two-dimensional parameters are given by \cite{boal2000,LL7}
\beq
Y = Y_3 h ~~~ {\rm and} ~~~  \kappa = \frac{Y_3 h^3}{12(1-\sigma^2)}.
\ee
Expressing eq. (\ref{contact_radius}) in these variables, we have for large $y
\gg y_{c}$
\beq
R^* \approx a_\infty \left(12(1-\sigma^2)y_c \right)^{1/4} \! \! \left(\frac{R_0^3}{Y_3 h^2} \right)^{1/2} \! \! (W-W_c)^{1/2}
\label{contact_radius_unscaled}
\ee with \beq W_c \equiv \frac{Y_3 h^3}{6(1-\sigma^2)R_0^2} \ee for the
critical adhesion energy. The large $y$ limit is appropriate since $y =
12(1-\sigma^2) R_0^2/h^2 \simeq 10^6$ in these experiments.

First, we compare our result to capsules with fixed shell thickness $h = 25.4$
nm and varying capsule's radius $R_0$ and contact radius $R^*$. In Fig. 4, the
experimental data from Ref. (\onlinecite{elsner2004}) are shown. Using typical
experimental values for $R^{*}$ and $R_{0}$ we can estimate $w-w_{c}$ to be of
the order $100$. Thus the critical strength of adhesion can entirely be
neglected in eq. (\ref{contact_radius_unscaled}). Fitting $R^*$ as a function
of $R_0$ with $W_{c} = 0$ we extract for the combination of adhesion energy,
Young modulus and Poisson ratio the estimate $W \simeq Y_3/(1-\sigma^2)^{1/2}
\cdot (1.4 \pm 0.1) \cdot 10^{-12}$ m. Estimates of the adhesion
energy become possible using experimental values of the elastic constants. The
Young modulus of the shell material is $(294 \pm 30)$ MPa
\cite{mueller2005,elsner2006}. (The value given in Ref. (\onlinecite{elsner2004}) is
severely overestimated and therefore not used.) The exact value of the Poisson ratio is unknown
but it is usually between 1/3 and 1/2. We then get $0.39$ mJ/m$^{2}$ $\lsim W
\lsim 0.52$ mJ/m$^{2}$, choosing these ranges of parameters. The adhesion
energies obtained this way are of the same order as the estimates in Ref.
(\onlinecite{elsner2004}). In Fig. 4 the
capsule's radius varies between 7 and 18 $\mu$m. We can therefore estimate the
critical adhesion energy $W_{c}$ to be in the range from $3$ $\mu \mathrm{J /
  m^{2}}$ to $24$ $\mu \mathrm{J / m^{2}}$. In Fig. 4 both the fit with
$W_{c} = 0$ and the behaviour of $R^{*}$ vs. $R_{0}$ for finite $W_{c}$ are
shown for comparison.

Second, we compare our result with capsules with fixed radius $R_0 = 10$
$\mu$m and fit the contact radius $R^*$ as a function of the shell thickness
$h$ (data not shown). From these data we obtain $W \simeq
Y_3/(1-\sigma^2)^{1/2} \cdot (1.3 \pm 0.8) \cdot 10^{-12}$~m, in accordance
with the previous estimate.

\begin{figure}
\includegraphics[scale=0.9]{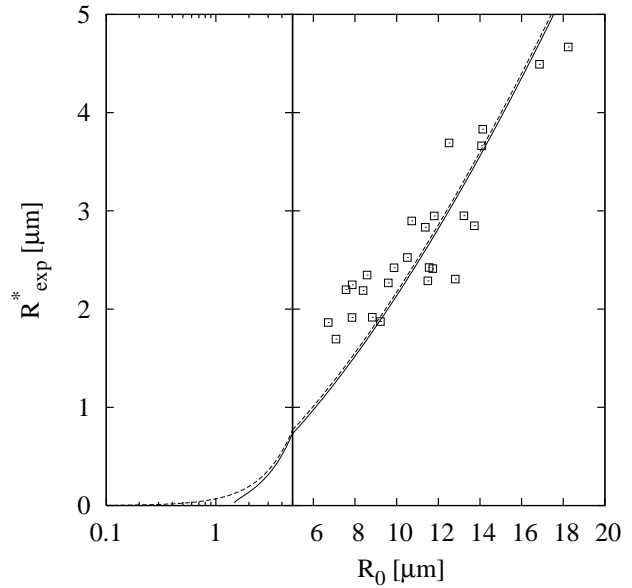}
  \caption{Comparison of experimental data ($\Box$, from Ref.
    (\onlinecite{elsner2004})) to the fit of eq. (\ref{contact_radius_unscaled}) with
    $W_{c} = 0$ (dashed line) and to the fit of eq. (\ref{contact_radius_unscaled}) for finite $W_{c}= 9$ $\mu \mathrm{J/m^{2}}$ as extracted from experimental data using the parameters $Y_3=294$
    MPa, $h=25.4$ nm, $\sigma=1/3$, and $R_0=10$ $\mu$m (full line). Note that the left part of this figure
    is scaled logarithmically to better show the influence of the critical
    adhesion energy $W_c$.}
\end{figure}

{\sl Summary. --} We have solved the shape equation numerically for elastic
microcapsules with finite shear elasticity adhering to a contact potential. We
have identified a threshold strength required for adhesion. For stronger
potentials the adhesion radius increases like a square root with an amplitude
depending on elastic constants which we here determined over the full range of
possible parameters.

{\sl Acknowledgement. --} Financial support of the DFG within the priority program SPP 1164 ``Nano- and microfluidics'' is gratefully acknowledged.

\end{document}